\input mn
\input psfig
\loadboldmathnames
\pageoffset{-2.5pc}{0pc}

\loadboldmathnames
\font\fivebmi=cmmib6
\font\sixbmi=cmmib6     \skewchar\sixbmi='177
\font\ninebmi=cmmib10 at 9pt    \skewchar\ninebmi='177
\newfam\bmifam
\textfont\bmifam=\ninebmi
\scriptfont\bmifam=\sixbmi
\scriptscriptfont\bmifam=\fivebmi

\long\def\hide#1{}

\mathchardef\alpha="710B
\mathchardef\beta="710C
\mathchardef\gamma="710D
\mathchardef\delta="710E
\mathchardef\epsilon="710F
\mathchardef\zeta="7110
\mathchardef\eta="7111
\mathchardef\theta="7112
\mathchardef\iota="7113
\mathchardef\kappa="7114
\mathchardef\lambda="7115
\mathchardef\mu="7116
\mathchardef\nu="7117
\mathchardef\xi="7118
\mathchardef\pi="7119
\mathchardef\rho="711A
\mathchardef\sigma="711B
\mathchardef\tau="711C
\mathchardef\upsilon="711D
\mathchardef\phi="711E
\mathchardef\chi="711F
\mathchardef\psi="7120
\mathchardef\omega="7121
\mathchardef\varepsilon="7122
\mathchardef\vartheta="7123
\mathchardef\varpi="7124
\mathchardef\varrho="7125
\mathchardef\varsigma="7126
\mathchardef\varphi="7127

\def\S{{\cal S}}\def\R{{\cal R}}\def\L{{\cal L}}
\def\pr{\mathop{\smash{\rm p}\vphantom{\sin}}}
\def\p{\partial}
\def\d{{\rm d}}

\def\lta{\mathrel{\spose{\lower 3pt\hbox{$\mathchar"218$}}
     \raise 2.0pt\hbox{$\mathchar"13C$}}}
\def\gta{\mathrel{\spose{\lower 3pt\hbox{$\mathchar"218$}}
     \raise 2.0pt\hbox{$\mathchar"13E$}}}
\def\spose#1{\hbox to 0pt{#1\hss}}

\newcount\eqnum \eqnum=1
\def\newe{\hbox{(\the\eqnum)}\global\advance\eqnum by1}
\def\eqname#1{\xdef#1{(\the\eqnum)}}
\def\ifundefined#1{%
   \expandafter\ifx\csname#1\endcsname\relax}
\def\refeq#1{\advance\eqnum by -#1\hbox{(\the\eqnum)}\advance\eqnum
by#1}
\def\ref#1{\ifundefined{#1} $\bullet$#1$\bullet$
           \else\hbox{\csname#1\endcsname}\fi}
\long\def\hide#1{}


\pagerange{530--536}    
\pubyear{1999}
\volume{302}

\begintopmatter  

\title{Kinematical signatures of hidden stellar discs}
\author{John Magorrian}
\affiliation{Canadian Institute for Theoretical Astrophysics,
University of Toronto, 60 St George Street, Toronto, Ontario, Canada M5S 3H8}

\shortauthor{S.J. Magorrian}
\shorttitle{Kinematical signatures of hidden stellar discs}


\abstract {The deprojection of the surface brightness distribution
of an axisymmetric galaxy does not have a unique solution unless the
galaxy is viewed precisely edge-on.  I present an algorithm that finds
the full range of smooth axisymmetric density distributions consistent
with a given surface brightness distribution and inclination angle,
and use it to investigate the effects of this non-uniqueness on the
line-of-sight velocity profiles (VPs) of two-integral models of both
real and toy discy galaxies viewed at a range of inclination angles.
Photometrically invisible face-on discs leave very clear signatures in
the minor-axis VPs of the models (Gauss--Hermite coefficients
$h_4\gta0.1$), provided the disc-to-bulge ratio is greater than about
$3\%$.  I discuss the implications of these hitherto neglected discs
for dynamical modelling.}


\maketitle  

\section{Introduction}

A fundamental problem in astronomy is determining the distribution of
instrinsic shapes of elliptical galaxies.  One way to do this is by
assuming that every galaxy is a triaxial ellipsoid, and then to use
simple photometrical and kinematical information to try to constrain
the distribution of axis ratios of their outer regions (e.g.,
Binney~1985; Franx, Illingworth \& de Zeeuw~1991; Statler~1994).
About a third of all ellipticals, however, have ``discy'' isophote
distortions (Bender et al.~1989), steep power-law central density
cusps, and show no obvious indications of triaxiality (Faber et
al.~1997; Kormendy \& Bender~1996; see also Merritt \& Quinlan~1998).
Little is known about the discs that most probably cause these
distortions.  For example, it is unclear whether discy ellipticals
form a continuous sequence in disc-to-bulge ratio with S0 galaxies
(Scorza \& Bender~1995).  Moreover, it is possible that quite large
discs could lurk in those power-law galaxies that do not have obvious
isophote distortions: Rix \& White (1990) have shown that even the
large discs in S0 galaxies are photometrically undetectable for about
50\% of orientation angles.  Therefore, rather than trying to
determine axis ratios for these power-law galaxies under the
assumption of ellipsoidal symmetry, it is perhaps more interesting to
use kinematical information to try to determine (or at least
constrain) their full central stellar density distributions
$\rho(R,z)$ under the assumption of axisymmetry.

The ease with which discs can be hidden can be understood using the
Fourier slice theorem (Rybicki~1987): in Fourier space, the surface
brightness distribution of an axisymmetric galaxy with a given
inclination angle~$i$ provides information about the density only
outside a ``cone of ignorance'' of opening angle $90^\circ-i$ around
the galaxy's symmetry axis, where $i=90^\circ$ for an edge-on galaxy.
Gerhard \& Binney (1996) have explicitly constructed a family of
disc-like ``konus densities'' whose Fourier transforms vanish outside
the cone of ignorance, and therefore whose densities project to zero
surface brightness.  Thus, the photometric deprojection problem is not
unique for non-edge-on galaxies: any solution $\rho(R,z)$ is
degenerate to the addition of disc-like konus densities.  Further
examples of konus densities are given by Kochanek \& Rybicki (1996)
and van den Bosch (1997).

Konus densities are quite difficult to construct analytically, so a
numerical approach is needed to explore the full range of possible
solutions to the photometric deprojection problem.  Such an approach
was taken by Romanowsky \& Kochanek (1997; hereafter RK) who used a
maximum penalized likelihood algorithm, penalizing their solutions
with ``bias functions'' that influenced the shape of the final
solution chosen, as well as providing the smoothness criterion that is
necessary for any discrete numerical approximation.  They found that
the uncertainties in the deprojection were even greater than one would
expect from the simple analytical examples of konus densities above.
RK also used Jeans equations modelling to predict the projected
second-order velocity moments corresponding to each solution, under
the assumption of a constant mass-to-light ratio and a two-integral
distribution function.  They found that the second-order moments were
only very weakly affected by the non-uniqueness of the deprojection:
different solutions gave almost identical projected second-moment
profiles.

Real kinematical measurements, however, include more information than
just the second-order velocity moments.  Indeed, nowadays one
routinely measures the full line-of-sight velocity profiles (VPs) of
galaxies.  In this paper, I use simple models to investigate whether
using VPs can place stronger constraints on the intrinsic shapes of
axisymmetric galaxies.  It is organized as follows.  Section~2
describes the deprojection problem in numerical terms and gives my
method of solution.  Like RK, I use maximum penalized likelihood, but
with a penalty function that trades off the elegance of their bias
functions for comprehensibility and ease of use.  I also use an
interpolation scheme that is better able to handle the cusped density
profiles present in real galaxies.  In Section~3, I demonstrate the
effects that konus densities have on the VPs of simple toy galaxy
models.  The discs in real galaxies are not strict konus densities, so
in Section~4 I use models of some real edge-on discy ellipticals to
investigate whether their VPs would contain any signature of the discs
were they viewed closer to face on.  Section~5 sums up and suggests
future work.

\section{Numerical solution of the photometric deprojection problem}

For projected quantities, let us choose a system of co-ordinates
$(x',y',z')$ with origin $O$ at the centre of the galaxy, the
$Oz'$-axis parallel to the line of sight to the galaxy, and the
$Ox'$-axis parallel to the galaxy's projected major (minor) axis for
an oblate (prolate) model.  Real galaxies are close to spheroidal, so
assume that we are given discrete measurements of the projected
surface brightness distribution $I(x',y')$ on an elliptical polar grid
$(m',\theta')$ with $n_{m'}\times n_{\theta'}$ vertices at
$$\eqalign{
   x'_{ij}=\phantom{q'}m_i'\cos\theta_j'\cr
   y'_{ij}=q'm_i'\sin\theta_j',\cr
}
\eqname\prjgrid\eqno\newe$$
where $m_i'$ runs logarithmically between the radii of the inner- and
outer-most isophotes, $\theta_j'$ is spaced linearly between $0$ and
$\pi/2$, and $q'$ ($1/q'$) is the mean projected axis ratio of the
galaxy assuming it is oblate (prolate).  Let $\S_{ij}\equiv\log
I(m'_i,\theta'_j)$ and let $\Delta\S_{ij}$ be the corresponding
measurement error (assumed Gaussian).

We seek a three-dimensional luminosity density $\rho(R,z)$, which,
when viewed at a given inclination angle~$i$, projects to an
acceptable fit to this $\S_{ij}$.  We expect this $\rho(R,z)$ to be
roughly spheroidal, and to have a radial profile that is locally
well-approximated by some power law.  Hence I represent $\rho(R,z)$ as
$\R(m,\theta)\equiv
\log\rho(m,\theta)$ on an $n_m\times n_\theta$ grid with vertices at
$$\eqalign{ R_{ij}=\phantom{q}m_i\cos\theta_j\cr z_{ij}=q
   m_i\sin\theta_j,\cr }
\eqname\intgrid\eqno\newe$$
where the intrinsic axis ratio $q$ is related to $q'$ by
$q'^2=q^2\sin^2i+\cos^2i$, $\theta_j$ runs linearly from $0$ to
$\pi/2$, and $m_i$ runs logarithmically from $m'_1$ to a few times
$m'_{n_m'}$.  Values of $\rho(m,\theta)$ between grid points are
obtained by interpolating linearly in $\R$ and exponentiating.  The
calculation of the model's projected surface brightness $\hat\S_{ij}$
on the grid~\ref{prjgrid} is outlined in the Appendix.

Given these $\hat\S_{ij}$, the goodness of fit is measured by
$$\chi^2=\sum_{i=1}^{n_{m'}}\sum_{j=1}^{n_{\theta'}}
\left( \S_{ij}-\hat\S_{ij}\over\Delta\S_{ij} \right)^2.
\eqno\newe$$
A reasonable model will have $\chi^2\simeq
n_{m'}n_{\theta'}\pm\sqrt{2n_{m'}n_{\theta'}}$.

There will be many distributions $\R_{ij}$ that will have a value of
$\chi^2$ in this range, but not all of them will be acceptably smooth.
In addition, one would like the ability to choose among the various
acceptable solutions, for example by preferring those with a given
degree of disciness or boxiness.  Given these reasonable prior
prejudices (let us denote them as $I$), and an observed $\S_{ij}$,
Bayes' theorem gives the relative plausibility of any set of $\R_{ij}$
as
$$\pr(\R\mid \S,I)\propto\pr(\S\mid\R,I)\pr(\R\mid I),
\eqno\newe$$
where the likelihood $\pr(\S\mid\R)=\exp(-{1\over2}\chi^2)$ and the
prior $\pr(\R\mid I)$ expresses our preconceptions (smoothness,
isocontour shape) about the form of $\rho(R,z)$.  Taking the logarithm
of equation~\refeq1\ yields an expression for the penalized
log-likelihood
$$\L\equiv-{1\over2}\chi^2+P_I[\R],
\eqno\newe$$
where the penalty function $P_I[\R]\equiv\log\pr(\R\mid I)$ penalizes
those solutions that do not fit our smoothness and isocontour-shape
requirements.

\subsection{The penalty function}

One of the simplest ways of penalizing non-smooth solutions is by
using the mean-square second derivative:
$$\eqalign{
P_{\rm smooth}[\R] & = {C_m\over n_\theta}
\sum_{i,j}
\left(\R_{i+1,j}-2\R_{ij}+\R_{i-1,j}\over\Delta \log m \right)^2\cr
&\quad + {C_\theta\over n_m}\sum_{i,j}
\left(\R_{i,j+1}-2\R_{ij}+\R_{i,j-1}\over\Delta\theta\right)^2,\cr
}
\eqname\smooth
\eqno\newe$$
where $C_m$ and $C_\theta$ are the weights given to radial and angular
smoothness and I have omitted some uninteresting constant factors.
Sensible values for the weights can be estimated as follows.  Consider
a typical galaxy with $\rho\sim m^{-4}$ in the outer parts, rolling
over to $\rho\sim m^{-1}$ in the centre.  Thus $\p\R/\p\log m$ changes
from $-4$ in the outer parts to $-1$ near the centre, and so the first
sum in equation~\refeq1\ must be at least $3^2/n_m$.  Choosing
$C_m=-(2n_{m'}n_{\theta'})^{1/2}n_m/\lambda_m^2$, an increase in the
RMS change in $\p\R/\p\log m$ by an amount~$\lambda_m$ is considered
as bad as an one-sigma increase in $\chi^2$ of
$(2n_{m'}n_{\theta'})^{1/2}$.  I choose $\lambda_m=3$ for all the
results in this paper.  Similarly, for the angular smoothing weight I
choose $C_\theta=-(2n_{m'}n_{\theta'})^{1/2}
n_\theta/\lambda_\theta^2$.  Since not much is known about galaxies'
angular profiles, it is not so clear how to choose $\lambda_\theta$ a
priori, but experiments with real galaxies show that
$\lambda_\theta\simeq0.5$ is about right.

The shape (degree of disciness or boxiness) of $\rho(m,\theta)$ can be
measured by its $\cos4\theta$ Fourier coefficients, defined as
$$c_4(m)\equiv\int_{-\pi/2}^{\pi/2} \rho(m,\theta)\cos4\theta\,\d\theta,
\eqno\newe$$
similar to the various definitions often used to quantify isophote
shapes (e.g., Jedrzejewski 1987; Bender \& M\"ollenhoff 1987).  At
radii where the model is locally discy (boxy) $c_4$ will be positive
(negative).  The penalty function
$$P_{\rm shape}[\R]=\left(n_{m'}n_{\theta'}\over n_m\right)^{1/2}\sum_i
\left(c_4(m_i)-d\over\Delta d\right)^2,
\eqname\shape\eqno\newe
$$
favours solutions with $c_4\simeq d\pm\Delta d$, provided the
observations do not place strong constraints on $c_4$.  If on the
other hand the observations constrain $c_4$ to lie outside the range
$d\pm\Delta d$, $P_{\rm shape}$ biases the solution towards $c_4=d$,
with the parameter $\Delta d$ controlling the strength of the bias.

A final, not-so-obvious, necessary constraint on the shape of the
density isocontours is that $\p\R/\p\theta$ be non-negative at
$\theta=0$ and $\pi/2$.  I use
$$\eqalign{
P_{\rm nn} & = {C_{\rm nn}\over n_m}
 \sum_i \big(\max(0,\R_{i,2}-\R_{i,1})\big)^2 +\cr
& \qquad{C_{\rm nn}\over n_m}
 \sum_i\big(\max(0,\R_{i,n_\theta}-\R_{i,n_\theta-1})\big)^2,\cr
}
\eqname\nonneg\eqno\newe
$$
to penalize each radial grid point that does not satisfy these
conditions by an amount $C_{\rm nn}=-4\sqrt{n_{m'}n_{\theta'}}$ .

Taking equations~\ref{smooth}, \ref{shape} and \ref{nonneg} together,
the full penalty function is $P_I=P_{\rm smooth}+P_{\rm shape}+P_{\rm
nn}$.

\subsection{Finding the best solution}

The following procedure is used to find the density $\rho(m,\theta)$
that maximizes the penalized log-likelihood $\L$.  There may well be
more efficient schemes, but this one is easy to implement and works
well.
First the parameters $(L,m_0,\alpha,\beta)$ of
the distribution 
$$\rho(m)={L\over4\pi qB(3-\alpha,3-\beta)}{m_0^{\beta-3}\over
m^\alpha(m_0+m)^{\beta-\alpha}},
\eqname\inifit\eqno\newe$$
that minimize $\chi^2$ are found
using the Levenberg--Marquardt algorithm (Press et al., 1992).
The function $B(\cdot,\cdot)$ in~\refeq1\ is the usual Beta function.
Then the following simulated annealing scheme (Metropolis et al.,
1953; Press et al. 1992) is used to improve this initial
$\R(m,\theta)$:

\begingroup
\item{(1)} Calculate $\L[\R]$ for this initial guess.

\item{(2)} Make a change $\delta\R$ to $\R$.  Calculate
$\hat\S[\R+\delta\R]$, $P_I[\R+\delta\R]$ and thus $\L[\R+\delta\R]$.

\item{(3)} If $\Delta\L\equiv\L[\R+\delta\R]-\L[\R]>0$ accept the
change $\delta\R$; otherwise accept it with probability
$\exp(\Delta\L/T)$, where the ``temperature'' $T$ is defined below.

\item{(4)} Go back to step 2.

\endgroup

In this scheme each grid point $(i,j)$ has an associated
$\Delta\R_{ij}$ that governs how much $\R_{ij}$ can change by in step
(2).  The natural initial value for all the $\Delta\R_{ij}$ is the RMS
difference between $\S_{ij}$ and the $\hat\S_{ij}$ of
equation~\ref{inifit}.
In step (2), each change $\delta\R$ is made by choosing one grid point
$(i,j)$ at random and adding $r\Delta\R_{ij}$ to $\R_{ij}$, where $r$
is a random number uniformly distributed in $(-1,1)$.
Whenever a change $\delta\R_{ij}$ is accepted (step~3),
$\Delta\R_{ij}$ is increased by a small factor (say 1.2); otherwise
$\Delta\R_{ij}$ is decreased by the same factor.  This ensures that
the changes the program makes are as large as possible.

Given the initial $\delta\R_{ij}$, a few times $n_mn_\theta$
iterations of step (2) are made, and the initial temperature $T$ is
chosen to be the mean value of $|\Delta\L|$.  This ensures that
roughly half the steps will be accepted.
Every $8n_mn_\theta$ iterations the temperature is set equal to the
mean change in $|\Delta\L|$ of all the accepted steps in the previous
$8n_mn_\theta$ iterations.  The program stops with this temperature is
less than $\sqrt{n_{m'}n_{\theta'}}$.

\section {Kinematical effects of konus densities}

\beginfigure{1}
\centerline{\psfig{file=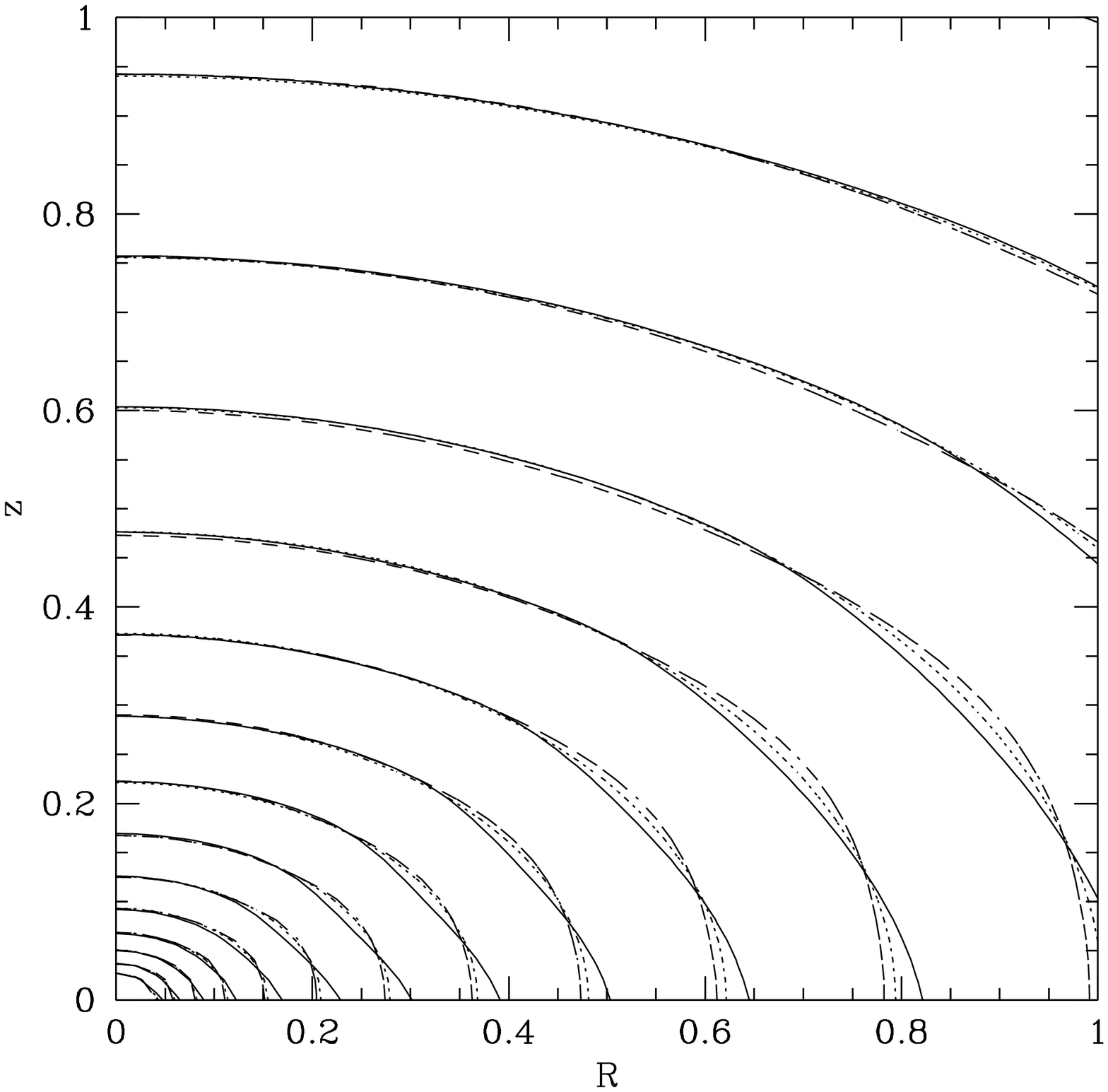,width=\hsize}}
\caption{{\bf Figure 1.} The most discy
(solid contours) and most boxy (dashed contours) density distributions
$\rho(R,z)$ that project to the surface brightness distribution of a
flattened Jaffe model viewed at $i=60^\circ$.  The dotted contours
plot the density of the original Jaffe model.}
\endfigure
\noindent
Before investigating real galaxies, it is worthwhile pausing to look
at the effects that konus densities can have on simple toy galaxies.
I use a Jaffe (1983) model (equation~\ref{inifit} with $\alpha=2$ and
$\beta=4$) with unit scale radius, unit luminosity and intrinsic axis
ratio $q=0.6$, viewed quite close to edge-on at $i=60^\circ$.  The
projected surface brightness distribution is placed on the
grid~\ref{prjgrid} with $n_{m'}\times n_{\theta'}=40\times7$, and
$m'_1=0.01$ and $m'_{40}=10$.  I seek the luminosity density
$\rho(R,z)$ on grid~\ref{intgrid} with $n_m\times n_\theta=60\times10$
vertices with $m_i$ running from $0.01$ to $100$.  As a test of the
accuracy of the numerical projection, placing the exact $\rho(R,z)$ on
this grid and projecting yields an $\hat\S_{ij}$ that has an RMS
difference of only 0.05\% from the results obtained by Jaffe's exact
expression.  This is much smaller than typical observational errors.

\beginfigure{2}
\centerline{\psfig{file=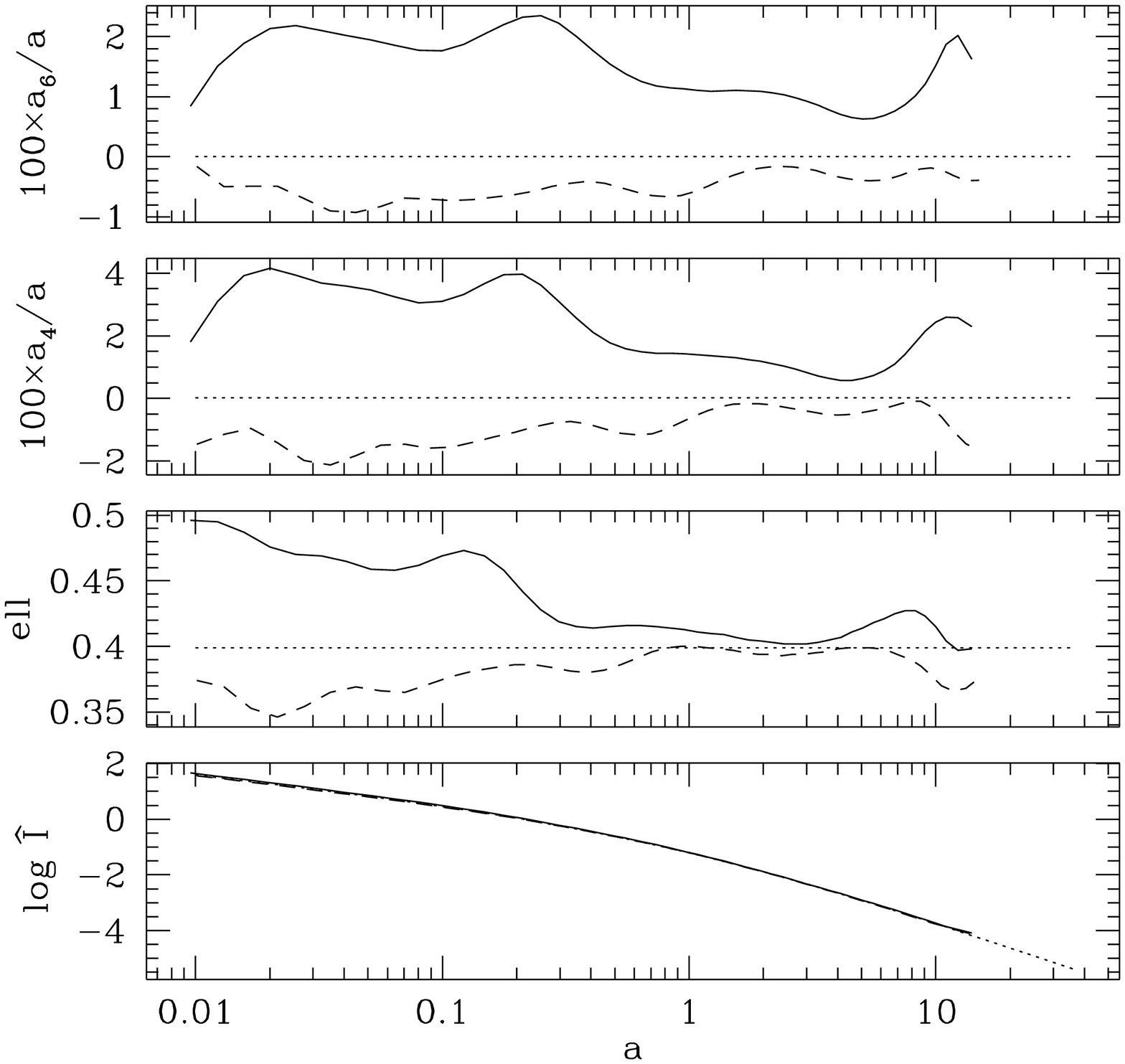,width=\hsize}}
\caption{{\bf Figure 2.} The surface brightness, ellipticity and
isophotal shape profiles of the models in Fig.~1 viewed edge-on.  I
use Bender \& M\"ollenhoff's (1987) definition of the isophotal shape
parameters $a_n$.}
\endfigure

To stop the deprojection program finding the exact solution
immediately, I fix $\alpha=1.5$ in the fitting of the initial guess
(equation~\ref{inifit}).  This initial guess has $L=0.49$, $m_0=0.26$,
$\beta=3.6$ and an RMS error of~0.12.  With $\Delta\S_{ij}=0.5\%$,
$\Delta d=0.03$, and $d=+0.1$ (discy bias) or $-0.1$ (boxy bias), the
program stops after about 70000 iterations of the Metropolis algorithm
have reduced this RMS error to around $0.56\%$.  To look for
systematic deviations between $\S_{ij}$ and $\hat\S_{ij}$, I use
Bender \& M\"ollenhoff's (1987) method to analyze the isophotes of the
solutions.  This yields the semi-major axis and ellipticity
$(a,\epsilon)$ of the best-fit ellipse to each isophote, along with
isophotal shape parameters $a_4,a_6,\ldots$, obtained by expanding the
angular dependence of the radial deviation between this best-fit
ellipse and the isophote as a Fourier series.  Both the boxy and discy
solutions above have $a_4/a\simeq\pm0.4\%$, which would be just on the
threshold of detectability for real observations, implying that
neither solution is acceptable.

Letting the program continue with $\Delta\S_{ij}=0.05\%$ in each case,
it stops with an RMS error of 0.03\% (smaller than the 0.05\%
numerical accuracy) after a further $200000$ or so iterations.  The
resulting density distributions are plotted on Fig.~1.
The projection of these densities yields isophotes with
$|a_{2n}/a|<0.03\%$, perfectly elliptical for all practical purposes.
Increasing $|d|$ and decreasing $\Delta d$ does not change the
solutions significantly, indicating that they are the most extreme
given the smoothness constraints.

These densities project to the surface brightness profile of a Jaffe
model for any $i<60^\circ$.  So, for a fraction $\cos i=50\%$ of
random orientations there is no sign of deviations from pure
ellipticity in the models' isophotes.  But when the models are viewed
edge-on, the disciness or boxiness becomes quite apparent (Fig.~2).

The konus density obtained by subtracting one of these solutions from
the other has almost exactly the same dependence on radius as the
original Jaffe model, except at the very centre where a konus density
can be no steeper than $r^{-1}$ (van den Bosch~1997).  Unlike the
analytical konus densities found by Gerhard \& Binney (1996), Kochanek
\& Rybicki (1996) and van den Bosch (1997), this numerical konus
density has only approximately a $\cos n\theta$ angular dependence and
there are no special angles $\theta$ along which its radial profile is
particularly shallow.

\beginfigure*{3}
\centerline{\psfig{file=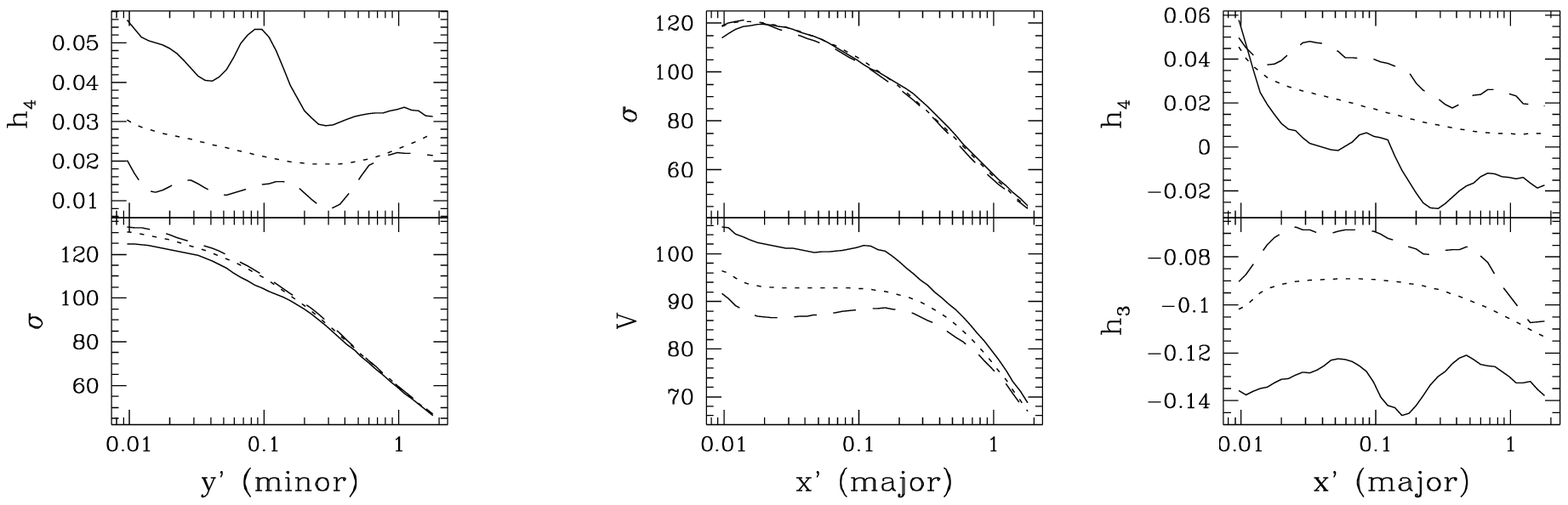,width=\hsize}}
\caption{{\bf Figure 3.} The major- and minor-axis
line-of-sight velocity profiles of the models in Fig.~1 (inclination
$i=60^\circ$), obtained by modelling each as a two-integral isotropic
rotator.  The panel on the left shows the Gauss--Hermite parameters of
the minor-axis VPs for the discy (solid curves), boxy (dashed curves)
and the original Jaffe (dotted curves) models.  The other two panels
show the Gauss--Hermite parameters of the major-axis VPs. }
\endfigure

To investigate the kinematical effects of the photometric degeneracy,
I calculate the isotropic-rotator VPs of each model using the
moment-based method of Magorrian \& Binney (1994).  It is convenient
to present the resulting VPs using the Gauss--Hermite parameterization
of van der Marel \& Franx (1993).  This gives the parameters $V$ and
$\sigma$ of the best-fitting Gaussian to each VP, along with
coefficients $h_3,h_4,\ldots,$ that describe how the VP deviates from
this Gaussian.  VPs with prograde wings steeper than retrograde ones
have $h_3<0$, while VPs more triangular (flat-topped) than the
Gaussian have $h_4>0$ ($h_4<0$).  Figure~3 shows the VPs of the discy
and boxy models in this form, along with the VPs of the original Jaffe
model.  On the major axis the discy model has relatively high $V$ and
negative $h_3$.  The slight bias towards circular orbits that causes
this also affects the discy model's minor-axis VPs, giving them
significant high-velocity wings and thus raising their $h_4$
coefficients.  (A stronger version of this phenomenon was first noted
by Scorza \& Bender (1995) in the minor-axis VPs of the discy
elliptical NGC~4660.)  The differences between the models'
Gauss--Hermite coefficients are approximately the same as the errors
in high signal-to-noise measurements of real VPs.  It is therefore
plausible that one could use VPs to constrain the internal $\rho(R,z)$
structure of a real galaxy with a somewhat larger disc than this
simple toy model.


\section{Kinematical effects of realistic discs}

To investigate the detectability of discs in real galaxies, I consider
what the nearby edge-on E7/S0 galaxy NGC~3115 looks like when viewed
at different orientations.  Scorza \& Bender (1995, hereafter SB) have
used a photometric disc--bulge decomposition to show that this galaxy
is consistent with a razor-thin disc viewed at $i=84^\circ$ embedded
in an almost perfectly elliptical bulge.  The $\rho(R,z)$ profiles I
obtain by deprojecting SB's photometry of this galaxy for $i=84^\circ$
and $i=90^\circ$ (Fig.~4) show that it can equally well be considered
as a reasonably fat disc-like component embedded in a slightly boxy
bulge-like component.  Since the galaxy is viewed close to edge-on, it
may be possible to distinguish between the thin- and fat-disc models
using the $a_8$ and higher-order isophote shape coefficients and a
great deal of care (or, perhaps more sensibly, by dropping the $a_n$
parameterization altogether).  In what follows, let us assume that
NGC~3115 is reasonably well described by my $i=84^\circ$ model, with a
constant mass-to-light ratio and an two-integral distribution
function.  None of these assumptions are strictly true, but this
simple model does serve to illustrate some important points.

\beginfigure*{4}
\centerline{\psfig{file=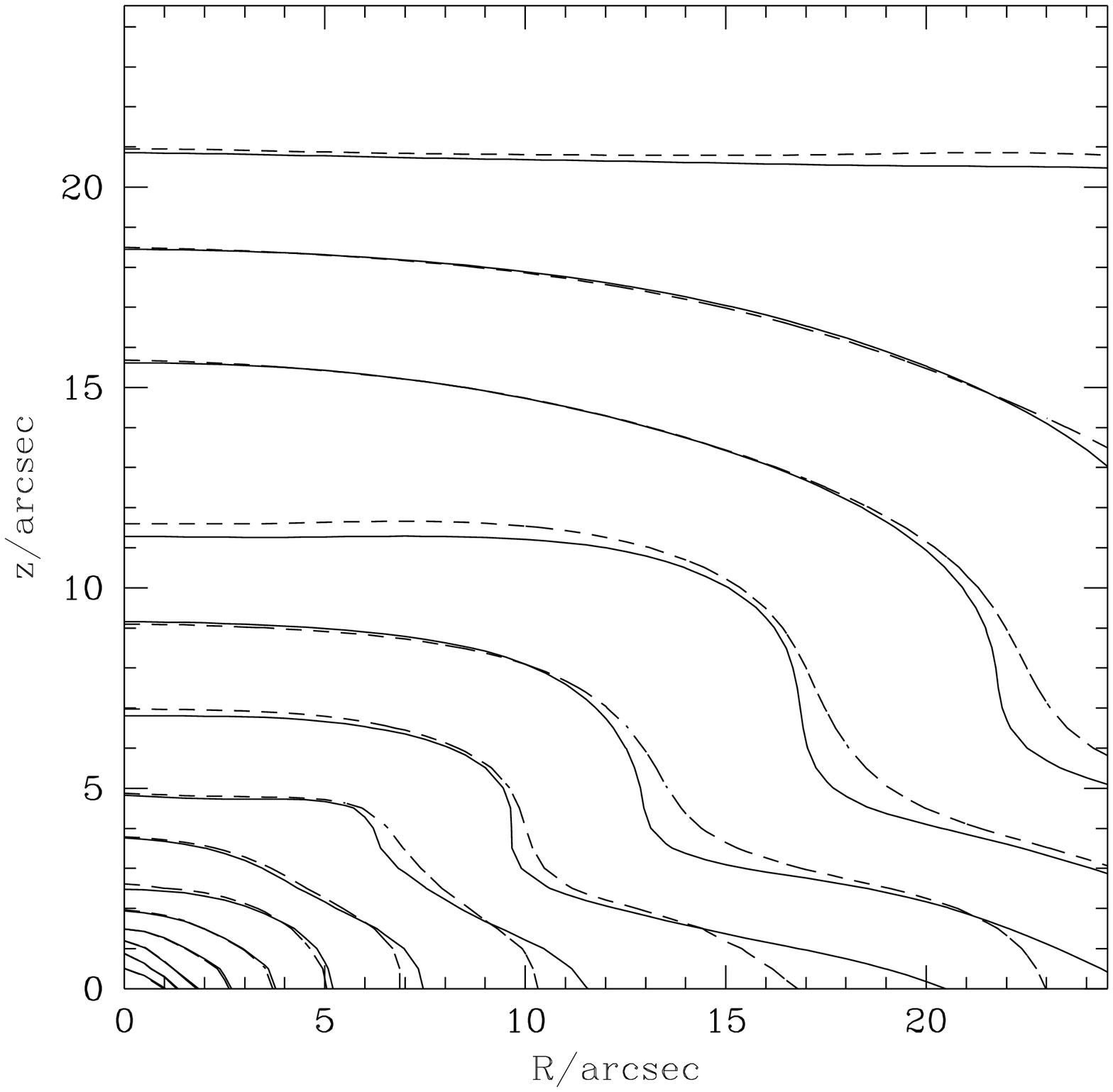,width=0.5\hsize}
            \psfig{file=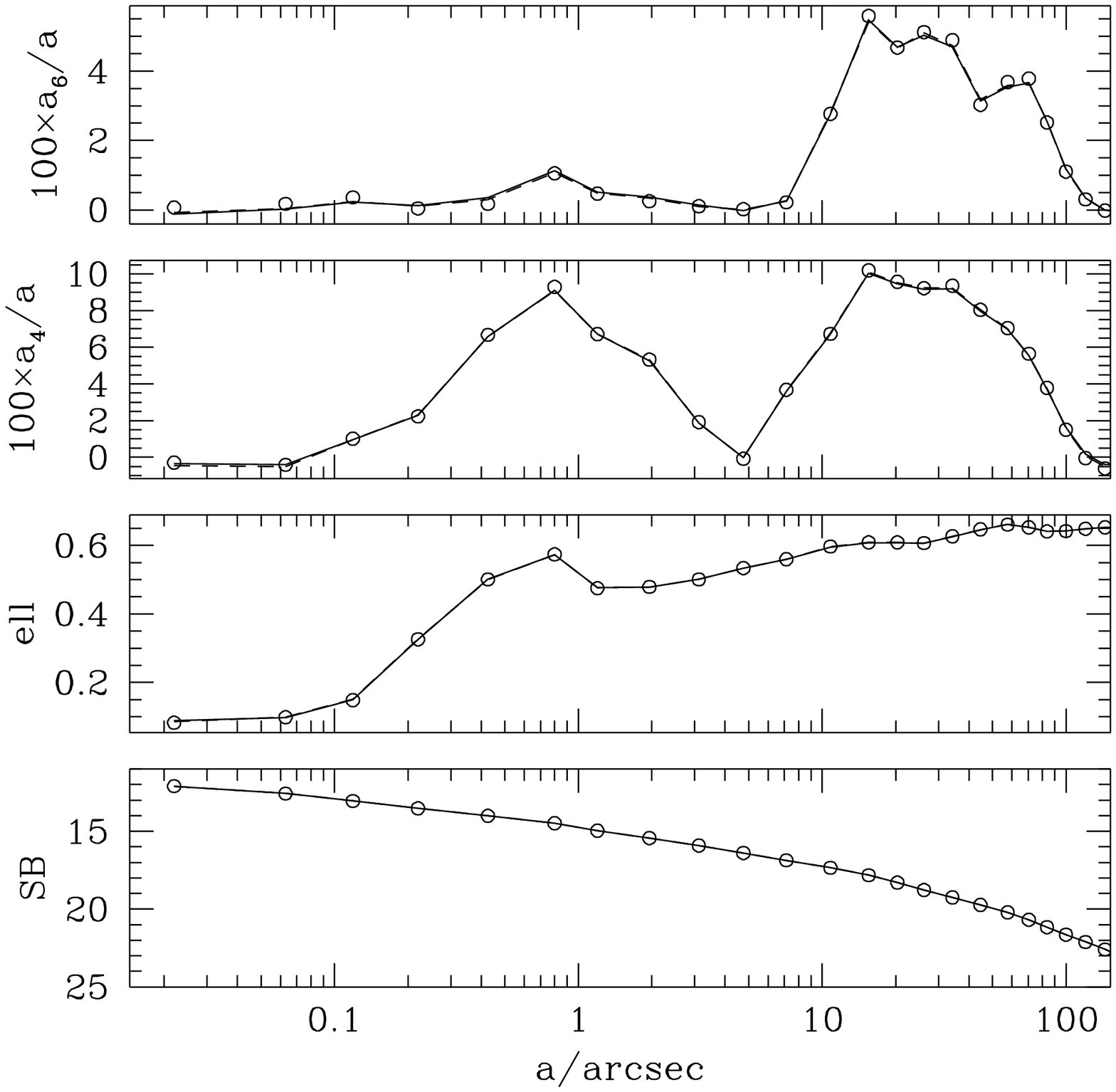,width=0.5\hsize}}
\caption{{\bf Figure 4.} (Left panel) Density distributions
$\rho(R,z)$ obtained for NGC~3115 assuming $i=84^\circ$ (solid
contours) and $i=90^\circ$ (dashed contours).  The panel on the right
shows that both models project to a reasonable fit to the observed
photometry (plotted as the circles).}
\endfigure

\beginfigure{5}
\centerline{\psfig{file=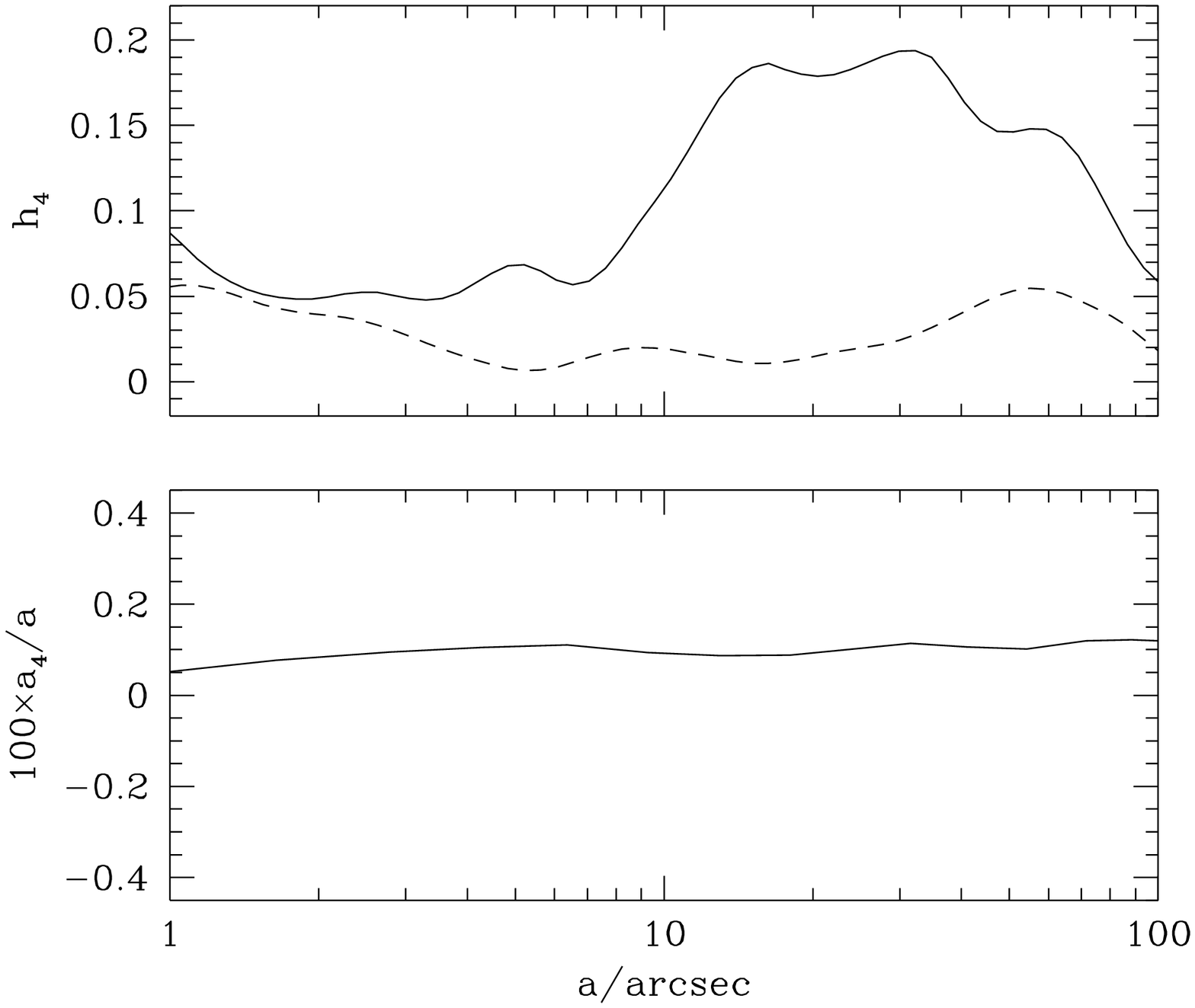,width=\hsize}}
\caption{{\bf Figure 5.} Evidence of the disc in
NGC~3115 when viewed almost face-on at $i=30^\circ$.  The small values
of the isophotal shape parameters $a_4/a$ on the bottom panel show
that there is no photometric sign of the disc at this orientation.  On
the other hand, the disc leaves a strong signature in the $h_4$
coefficients of the galaxy's minor-axis VPs (solid curve in top
panel).  For comparison, the minor-axis $h_4$ coefficients of the most
closely spheroidal model that fits the $i=30^\circ$ photometry are
plotted as the dashed curve.}
\endfigure

\beginfigure{6}
\centerline{\psfig{file=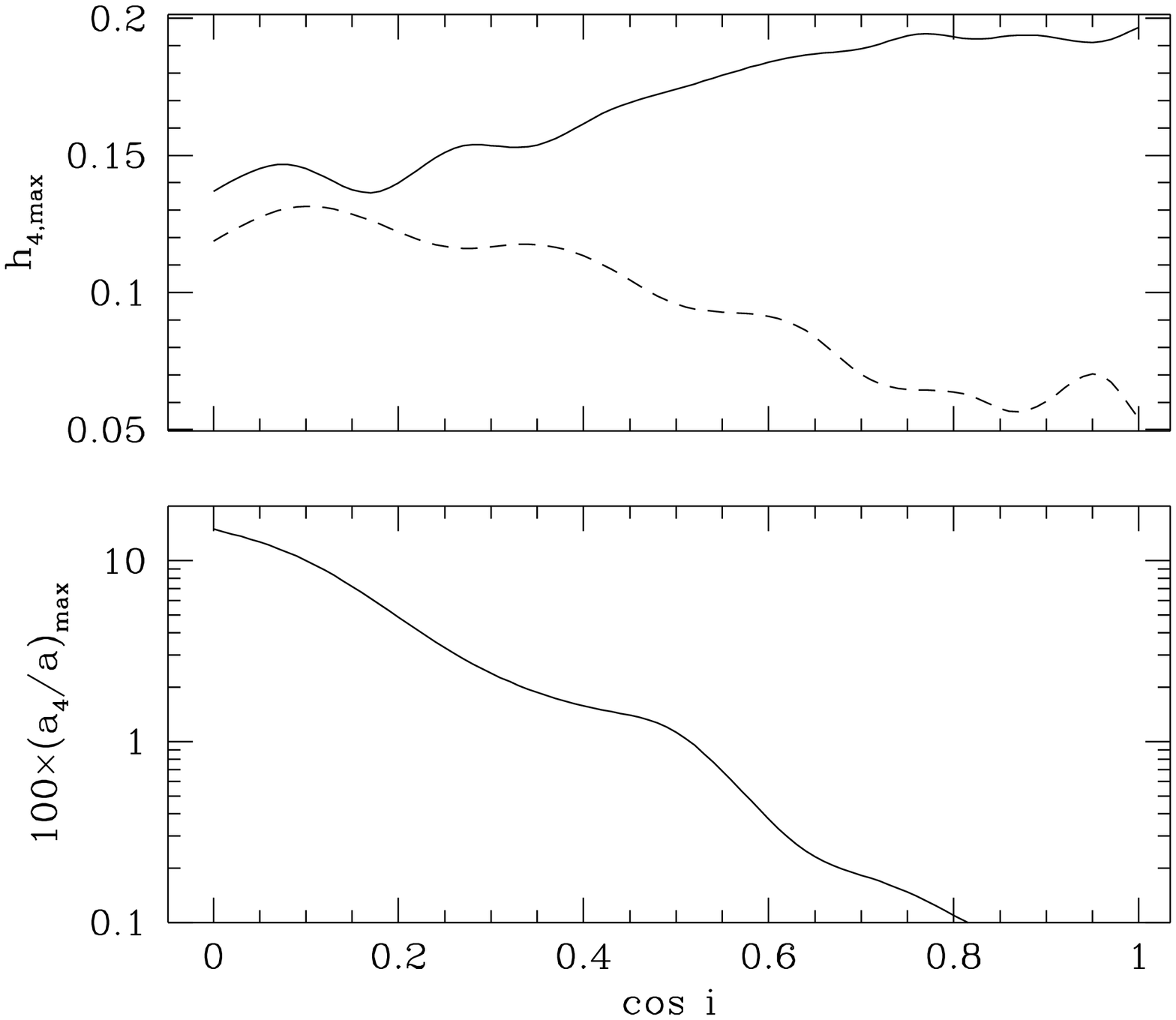,width=\hsize}}
\caption{{\bf Figure~6.} Detectability of the disc in NGC~3115 as a
function of inclination angle~$i$.  The maximum values of $a_4/a$ for
each inclination are plotted on the bottom panel, showing how quickly
the disc becomes photometrically invisible with decreasing
inclination.  The top panel shows the effect that the disc has on the
galaxy's maximum minor-axis $h_4$ coefficient (solid curve) for each
inclination.  For comparison, the maximum minor-axis $h_4$
coefficients of the most closely spheroidal two-integral model that fits
the observed photometry are plotted as the dashed curve.}
\endfigure

Suppose we view the galaxy close to face-on, at $i=30^\circ$.  Fig.~5
shows that in this case the maximum value of $a_4/a$ is only 0.2\% --
there is no indication of disciness in the photometry.  The disc does,
however, make itself evident in the model's minor-axis VPs, since the
extra circular orbits seen from above make the VPs more centrally
peaked, giving them large positive values of $h_4$ (solid curve in top
panel).  For comparison, deprojecting the $i=30^\circ$ surface
brightness distribution with penalty function $P_{\rm shape}$
parameters $d=0$ and $\Delta d=4(a_4/a)_{\rm max}=0.8\%$ yields the
model with the most closely spheroidal isodensity contours that fits
the data.  Because of the lack of a disc in this model, it has much
lower minor-axis $h_4$ coefficients (dashed curve).

The probability of observing a randomly oriented galaxy at inclination
angle~$i$ is proportional to $\cos i$.  On the lower panel of Fig.~6,
I plot the maximum $a_4/a$ of the model of NGC~3115 as a function of
$\cos i$, showing that for about 40\% of all randomly chosen
orientations $(a_4/a)_{\rm max}\lta0.4\%$.  Rix \& White (1990) have
found similar results using razor-thin exponential discs embedded in
$R^{-1/4}$ spheroids.  As the disc becomes less evident in the
photometry, however, its effect on the minor-axis VPs increases (upper
panel of figure).  Indeed, there is no inclination angle for which the
disc would not leave either a photometric or kinematic signature.
These results are not confined to NGC~3115: I find similar results
using SB's photometry to model the discy elliptical galaxies NGC~3377
and NGC~4660.

Finally, to understand how the detectability of discs correlates with
the disc-to-bulge ratio, I embed discs with a range of luminosities in
the toy Jaffe model of the preceding section.  Each disc has an
exponential radial profile, with unit scale radius and an isothermal
${\rm sech}^2(z/2z_0)$ vertical distribution with scale height
$z_0=0.1$.  Fig.~7 shows that for all inclination angles the disc is
clearly evident in either the photometry or the kinematics, provided
the disc-to-bulge ratio is at least 3\%.

\beginfigure{7}
\centerline{\psfig{file=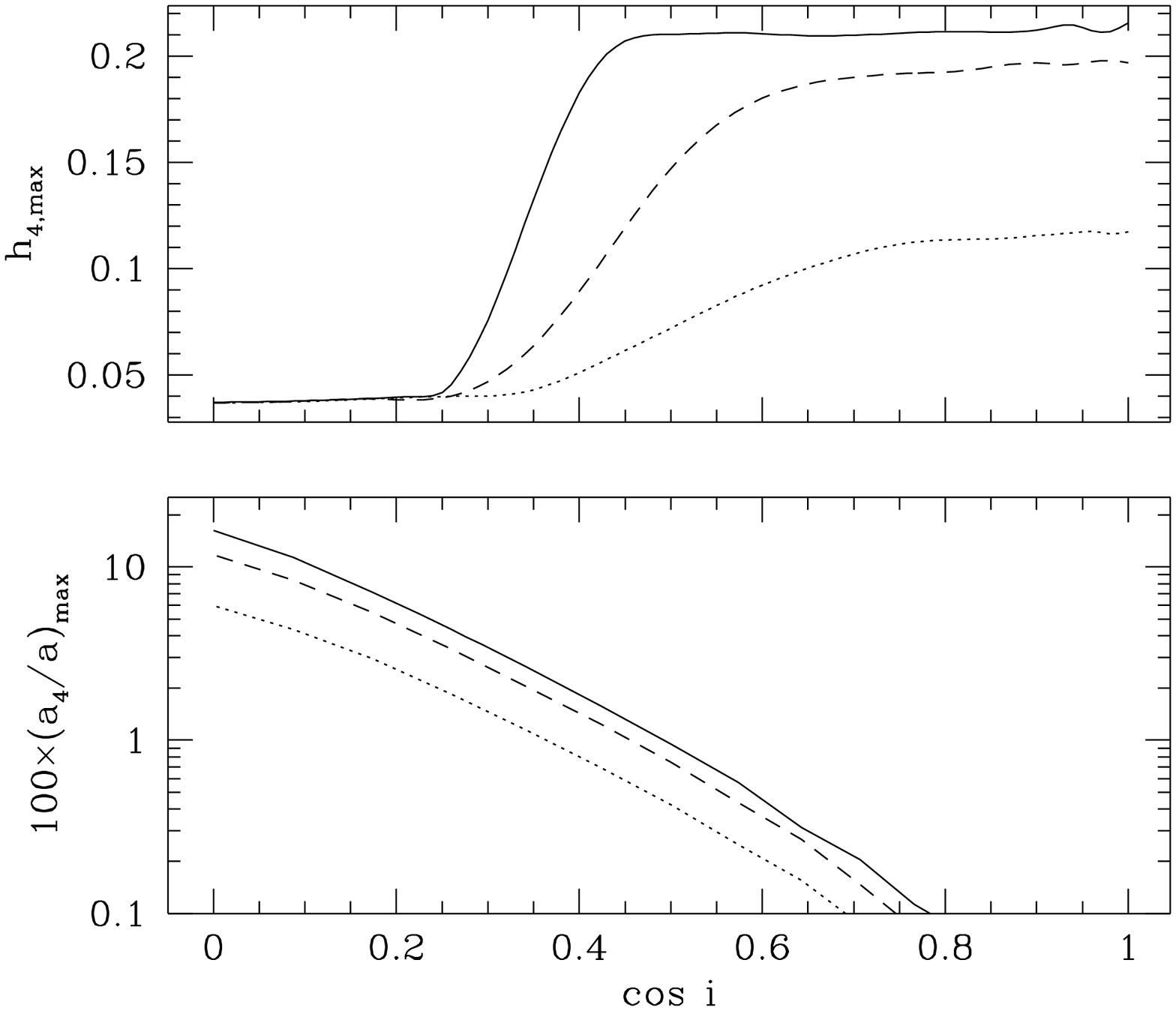,width=\hsize}}
\caption{{\bf Figure 7.} Detectability of the disc in the toy galaxy
model described in the text for disc-to-bulge ratios of 0.1 (solid
curves), 0.03 (dashed curves) and 0.01 (dotted curves).  As in 
Fig.~6, the bottom panel shows the maximum $a_4/a$ coefficient as a
function of inclination, while the top shows the maximum minor-axis
$h_4$ coefficient.}
\endfigure

\section {Conclusions}

It has been known for some time (Rix \& White~1990) that it is
possible that quite large, almost face-on discs could lurk in
elliptical galaxies without leaving any detectable signature in the
galaxies' photometry.  More sensitive photometry is unlikely to
improve the detectability of these discs: Gerhard \& Binney (1996)
have shown that the deprojection of any axisymmetric galaxy that is
not viewed exactly edge-on is formally degenerate to the addition of
disc-like konus densities.  Using two-integral models of nearby edge-on
discy ellipticals, I have shown that hidden discs do, however, leave
very strong, easily detectable signatures in the galaxies' minor-axis
VPs, provided the disc-to-bulge ratio is greater than about 3\%.

These results are based on two-integral models, but they have
important implications for more realistic three-integral models.  The
most sophisticated current modelling machinery (e.g., van der Marel et
al.~1998; Gebhardt et al.~1998) uses an {\it assumed} (usually
spheroidal) $\rho(R,z)$ and corresponding self-consistent potential to
constrain a galaxy's orbit distribution from its VPs.  Given, say, an
isotropic galaxy with a weak, invisible face-on disc, it is possible
that such machinery could fit the positive $h_4$ coefficients of the
galaxy's minor-axis VPs using a spheroidal $\rho(R,z)$ with a radially
biased orbit distribution.  (In principle, a particularly cold disc
might leave an unambiguous signature on the VPs, but it is unlikely
that this signature would survive the instrumental broadening that
real measurements suffer from.)  Thus it is important that some
attention be focused on modelling galaxies that are known to be close
to edge-on (i.e., are discy), so that the uncertainty in the
photometric deprojection is minimized.

Ultimately, however, one is interested in all the other galaxies whose
inclination angles, density distributions and orbit distributions are
unknown.  Until more general modelling machinery that ``knows'' about
the orbital structure of edge-on galaxies becomes available, it might
be worthwhile relaxing the self-consistency requirement in current
models and have them fit to VPs and {\it photometry}, rather than
fitting to an assumed, unobservable $\rho(R,z)$.  It would be possible
to iterate, using the potentials of the resulting density
distributions to generate proper self-consistent models.  This could
be an informative first step in trying to constrain both the internal
shapes and orbital structures of galaxies.

Unfortunately, it is not possible to test these ideas yet because
there are currently very few power-law galaxies with high
signal-to-noise measurements of minor-axis VPs.  Let us hope that this
is only a temporary problem.

\section*{Acknowledgments}
I thank Ralf Bender, James Binney, David Merritt and the referee,
Ortwin Gerhard, for enlightening discussions.  Financial support was
provided by NSERC.

\appendix
\section{The Projection along the line of sight}

The projected surface brightness distribution $\hat I(x',y')$ of the
density distribution $\rho(R,z)$ is 
$$\hat I(x',y')=\int_{-\infty}^\infty \rho(R,z) \,\d z'
\eqname\prjeq\eqno\newe$$
where $(R,z)$ is related to $(x',y',z')$ through the rotation
$$
\left(\matrix{x\cr y\cr z\cr}\right) =
\left(\matrix{1 & 0 & 0 \cr
              0 & \cos i & \sin i\cr
              0 & \sin i & -\cos i\cr}\right)
\left(\matrix{x'\cr y'\cr z'\cr}\right),
\eqname\xyz\eqno\newe$$
and $R^2=x^2+y^2$.  With the substitution
$$z'=\left(x'^2+y'^2\right)^{1/2}\sinh u,\eqname\subprj\eqno\newe
$$
the integral in equation~\ref{prjeq} can be approximated using the
trapezium rule as
$$\hat I(x',y') \simeq (x'^2+y'^2)^{1/2}\,
\d u \sum_{k=0}^n \rho(R_k,z_k) \cosh u_k,
\eqno\newe$$
with $u_0$ and $u_n$ chosen such that the corresponding $z'$ lie just
outside the edges of the model density grid, and the intermediate
$u_k$ spaced linearly between them with $\d u\equiv (u_n-u_0)/n$.  The
points $R_k$ and $z_k$ are the intrinsic co-ordinates of the point
$(x',y',z'_k)$ in the $k^{\rm th}$ term in the sum.  To ensure fair
sampling of the grid points along the line of sight, I choose
$n=4n_a$.

Since $\rho(m,\theta)$ is obtained by linear interpolation in
$\R(m,\theta)\equiv\log\rho(a,\theta)$ this sum can be rewritten in
the form
$$\eqalign{\sum_{k=0}^n E_k\exp\big( & A_k\R_{a_k,b_k} + B_k\R_{a_k+1,b_k}
\cr
& \quad+ C_k\R_{a_k,b_k+1} + D_k\R_{a_k+1,b_k+1}\big),\cr}
\eqno\newe$$
where the $(x',y')$-dependent coefficients $A_k\ldots E_k$, $a_k$ and
$b_k$ do not depend on $\R$ and so only have to be evaluated once for
each $(x'_{ij},y'_{ij})$.  Expressions for these constants are quite
easy to work out, but tedious to write out.

If, like Romanowsky \& Kochanek (1997), I had chosen to interpolate
linearly in $\rho$ rather than $\log\rho$, then equation~\refeq1\
would be replaced by an even simpler expression.  However, this
interpolation-in-$\rho$ scheme is unable to represent steeply cusped
density distributions accurately: using it, the RMS error in the
numerical projection of the toy model used in Section~3 is 1.6\%,
rather than the 0.05\% obtained with interpolation in $\log\rho$.  On
doubling $n_m$ to increase the resolution of the radial $m_i$ grid,
the error goes down only slightly to 1.3\%.  (On the other hand, the
error in the $\log\rho$ scheme goes down to 0.015\%.)  Thus, the extra
expense that interpolating in $\log\rho$ brings to the numerical
projection equation~\refeq1\ is easily justifiable.

\hide{It would be straightforward, though tedious, to include convolution
with a point-spread function in the calculation of $\hat\S_{ij}$.
}


\section*{References}
\beginrefs
\bibitem Bender R., M\"ollenhoff C., 1987, A\&A, 177, 71
\bibitem Bender R., Surma P., D\"obereiner S., M\"ollenhoff C.,
Madejsky R., 1989, A\&A, 217, 35
\bibitem Bender R., Saglia R.P., Gerhard O.E., 1994, MNRAS, 269, 785
\bibitem Binney J., 1985, MNRAS, 212, 767
\bibitem Faber S.M., et al., 1997, AJ, 114, 1771
\bibitem Franx M., Illingworth G., de Zeeuw T., 1991, ApJ, 383, 112
\bibitem Gebhardt K., et al., 1998, submitted to AJ
\bibitem Gerhard O.E., Binney J., 1996, MNRAS, 279, 993
\bibitem Jaffe W., 1983, MNRAS, 202, 995
\bibitem Jedrzejewski R.I., 1987, MNRAS, 226, 747
\bibitem Kochanek C.S., Rybicki G.B., 1996, MNRAS, 280, 1257
\bibitem Kormendy J., Bender R., 1996, ApJL, 464, 119
\bibitem Magorrian S.J., Binney J., 1994, MNRAS, 271, 949
\bibitem Merritt D., Quinlan G.D., 1998, ApJ 498 625
\bibitem Metropolis N., Rosenbluth A., Rosenbluth M., Teller A.,
Teller E., 1953, J. Chem. Phys., 21, 1087
\bibitem Press W.H., Flannery B.P., Teukolsky S.A., Vetterling W.T.,
1992, Numerical Recipes in C, 2nd edn.  Cambridge Univ. Press, Cambridge
\bibitem Rix H.-W., White S.D.M., 1990, ApJ, 362, 52
\bibitem Romanowsky A.J., Kochanek C.S., 1997, MNRAS, 287, 35
\bibitem Rybicki G., 1987, in de Zeeuw T., ed., Proc. IAU Symp. 127,
Structure and dynamics of Elliptical Galaxies. Reidel, Dordrecht, p. 397
\bibitem Scorza C., Bender R., 1995, A\&A, 293, 20 (SB)
\bibitem Statler T., 1994, ApJ, 425, 500
\bibitem van den Bosch F., 1997, MNRAS, 287, 543
\bibitem van der Marel R.P., Franx M., 1993, ApJ, 407, 525
\bibitem van der Marel R.P., Cretton N., de Zeeuw P.T., Rix H.-W.,
1998, ApJ, 493, 613
\endrefs
\bye